\documentclass{article}

\usepackage[preprint]{neurips_2025}

\usepackage[utf8]{inputenc} 
\usepackage[T1]{fontenc}    
\usepackage{hyperref}       
\usepackage{url}            
\usepackage{booktabs}       
\usepackage{amsfonts}       
\usepackage{nicefrac}       
\usepackage{microtype}      
\usepackage{xcolor}         
\usepackage{xspace}
\usepackage{graphicx}
\usepackage{wrapfig}
\usepackage{amsmath}
\usepackage{amssymb}
\usepackage{mathtools}
\usepackage{amsmath}
\usepackage{amssymb}
\usepackage{graphicx}
\usepackage{listings}

\usepackage{pifont}
\newcommand{\cmark}{\textcolor{green!70!black}{\ding{51}}}%
\newcommand{\xmark}{\textcolor{red}{\ding{55}}}%
\newcommand{\gray}[1]{\textcolor{gray}{#1}}
\usepackage{booktabs}
\usepackage{tabularx}

\lstdefinestyle{json}{
    basicstyle=\ttfamily\small,
    numbers=none,
    breaklines=true,
    morestring=[b]",  
    stringstyle=\color{green!50!black},
    showstringspaces=false,
    tabsize=2,
    frame=single,
    backgroundcolor=\color{gray!5},
    moredelim=**[is][\color{blue}]{@}{@},  
    moredelim=**[is][\color{red}]{|}{|},   
    captionpos=b
}

\lstdefinestyle{docstring}{
    basicstyle=\ttfamily\small,
    breaklines=true,
    frame=single,
    backgroundcolor=\color{gray!5},
    numbers=none,
    columns=flexible,
    keepspaces=true
}

\newcommand{\bluetext}[1]{\textbf{\textit{\textcolor[HTML]{4285F4}{#1}}}}
\newcommand{\orangetext}[1]{\textbf{\textit{\textcolor[HTML]{FF7B1A}{#1}}}}

\title{TalkPlay-Tools: Conversational Music Recommendation with LLM Tool Calling} 
\author{%
  Seungheon Doh~$^{1,2,}$\thanks{Equal contribution. Corresponding authors: seungheondoh@kaist.ac.kr, juhan.nam@kaist.ac.kr;}~,\quad Keunwoo Choi~$^{1,2,*}$,\quad Juhan Nam~$^{1}$ \\
  $^{1}$ KAIST, South Korea, \quad $^{2}$ talkpl.ai, USA \\ \color{magenta}{\url{https://talkpl-ai.github.io/talkplay_tools}}
}

\begin{document}

\maketitle

\begin{abstract}
While the recent developments in large language models (LLMs) have successfully enabled generative recommenders with natural language interactions, their recommendation behavior is limited, leaving other simpler yet crucial components such as metadata or attribute filtering underutilized in the system. We propose an LLM-based music recommendation system with tool calling to serve as a unified retrieval-reranking pipeline. Our system positions an LLM as an end-to-end recommendation system that interprets user intent, plans tool invocations, and orchestrates specialized components—boolean filters (SQL), sparse retrieval (BM25), dense retrieval (embedding similarity), and generative retrieval (semantic IDs). Through tool planning, the system predicts which types of tools to use, their execution order, and the arguments needed to find music matching user preferences, supporting diverse modalities while seamlessly integrating multiple database filtering methods. We demonstrate that this unified tool-calling framework achieves competitive performance across diverse recommendation scenarios by selectively employing appropriate retrieval methods based on user queries, envisioning a new paradigm for conversational music recommendation systems.
\end{abstract}

\section{Introduction}
\label{intro}

Music recommendation has long been shaped by how to filter databases according to user preferences. Early systems relied on boolean filtering~\cite{celma2010music} over listening history logs and metadata such as title, artist, and release year. The advent of large-scale music item catalogs and user-item feature learning algorithms~\cite{koren2009matrix} shifted the field toward retrieval and reranking pipelines driven by large-scale user and item embeddings~\cite{su2009survey, hu2008collaborative}. With advances in representation learning, the field has progressed beyond listening history to develop deeper content understanding~\cite{van2013deep, oramas2017deep}. Moreover, music representations for audio~\cite{lee2020disentangled}, lyrics~\cite{patra2017retrieving}, and visual artwork~\cite{hsia2018representation} has enabled example-based recommendation, while multimodal music representations~\cite{manco2022contrastive, huang2022mulan, doh2023toward, doh2024enriching, wu2023large, wu2025clamp, oramas2024talking, surana2025musicrs} has supported natural language query understanding. Recent advances in large language model (LLM)-based recommendation systems~\cite{doh2025talkplay, palumbo2025text2tracks} demonstrate that items can be represented in a semantic space and quantized into discrete Semantic IDs~\cite{rajput2023recommender, mei2025semantic}, enabling both generative recommendation and interactive dialogue with users. In these methods, users are allowed to engage actively to discover music that matches their goals and preferences through multi-turn interactions.

\begin{figure*}[h]
\centering
\includegraphics[width= \linewidth]{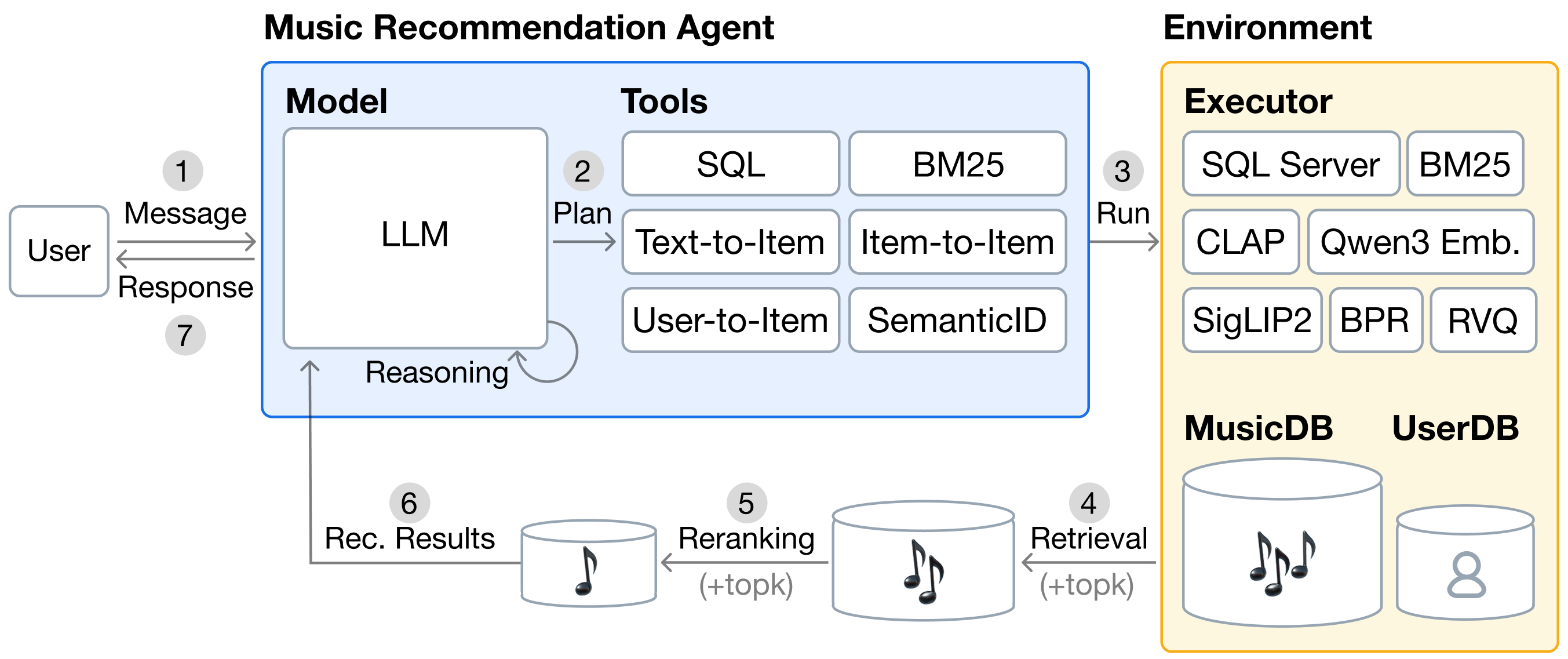}
\vspace{-5mm}
\caption{Overview of Music Recommendation Agents with Tool Calling.}
\vspace{-4mm}
\label{fig:main-overview}
\end{figure*}

However, relying on a single retrieval method has clear limitations in identifying items that satisfy user needs~\cite{weller2025theoretical}. Production-level recommenders~\cite{covington2016deep, jacobson2016music} operate based on multiple stages and routed retrieval–reranking pipelines and must strictly satisfy operational constraints (e.g., user profile, genre, mood, activity, newness, etc.), while simultaneously reflecting a listener's history and text queries. Without combining multiple types of retrieval methods, it cannot fully capture all the relevant context of recommendations.

In this paper, we present a conversational music recommender with tool-calling. Our contributions are: (i) formulating conversational music recommendation with multiple retrieval methods, enabling the LLM to predict tool types, order, and arguments conditioned on a user profile and a dialogue context; (ii) designing a unified tool set 
that composes of boolean filtering (SQL), sparse retrieval (BM25), dense retrieval across modalities (text-to-item, item-to-item, and user-to-item), and generative retrieval (Semantic IDs) under a single agent; and (iii) demonstrating zero-shot effectiveness on a conversational recommendation benchmark with improved Hit@K over strong baselines, together with detailed analyses of their behaviors. 


\section{Music Recommendation with Tool Calling}
\label{method}

Figure~\ref{fig:main-overview} illustrates our framework that consists of two main components. The first component is the \textit{Music Recommendation Agent}, comprising the LLM and tools. The second component is the \textit{External Environment} that executes the tools and performs the final recommendation through retrieval and reranking.

\subsection{Problem Formulation}
Given a user \(u\) with profile \(p_u\), a previous conversation state \(s_{t-1}\), and the user query \(q_t\) at the current turn \(t\), the proposed system produces a ranked list of music items \(m_t\) by (i) generating tool calls with a LLM and (ii) filtering database \(\mathcal{D}\) using the predicted tools \(\mathcal{C}_t\). Finally, the LLM is called again to generate a natural language response \(r_t\) that provides a conversational explanation of the recommendations while maintaining dialogue context.

\vspace{-4mm}
\begin{align}
\mathcal{C}_t &= \text{LLM}(q_t, s_{t-1}, p_u; \mathcal{P}_{tool}, \mathcal{T}) \\
m_t &= \text{ToolEnv}(\mathcal{C}_t; \mathcal{D}) \\
r_t &= \text{LLM}(m_t, q_t, s_{t-1}, p_u; \mathcal{P}_{response})
\end{align}
\vspace{-4mm}

Let \(\mathcal{T}\) denote the available tools, \(\mathcal{P}\) the prompt, and \(\mathcal{C}_t = \big[(\text{tool}_n, \text{args}_n)\big]_{n=1}^{N}\) the LLM-predicted sequence of tool calls with the number of tools \(N\). The tool environment executes each and every \((\text{tool}_n, \text{args}_n)\) pair to retrieve and rerank results into a final ranked list \(m_t\).

The tool execution environment operates as a sequential pipeline where each tool's output directly influences the subsequent tool's input space, i.e., each tool filters and refines the track pool for downstream tools. Therefore, the order of the tools significantly affects the final recommendation quality. Through the prompt $\mathcal{P}_{tool}$, we guide the LLM to perform recommendations not as a single operation, but as a staged process consisting of retrieval and reranking phases, ensuring that the tool execution follows a sequential pipeline.


\subsection{Tools}
Our framework incorporates diverse tools that enable precise and flexible music recommendation through structured filtering and semantic retrieval approaches. Table \ref{tab:tool_comparison} summarizes their capabilities, environments, and usage examples.

\begin{table}[!t]
\caption{Comparison of capabilities across different tools. Gray text indicates tools activated by in-context information rather than direct natural language user queries.}
\label{tab:tool_comparison}
\centering
\resizebox{\columnwidth}{!}{%
\begin{tabular}{llll}
\toprule
Tools & Capabilities & Environments & Query Examples \\ \midrule
\multicolumn{4}{l}{{\color[HTML]{9B9B9B} \textit{Boolean Retrieval}}} \\
SQL & Numeric Filtering & SQL Server & ``Recent songs over 130 BPM" \\ \midrule
\multicolumn{4}{l}{{\color[HTML]{9B9B9B} \textit{Sparse Retrieval}}} \\
BM25 & Lexical Matching & BM25 Index & ``Songs from Adele’s 21" \\ \midrule
\multicolumn{4}{l}{{\color[HTML]{9B9B9B} \textit{Dense Retrieval}}} \\
Text-to-Item & Semantic Matching & Qwen3, CLAP, SigLIP2 & ``Play a calm piano piece" \\
Item-to-Item & Multimodal Matching & CLAP, SigLIP2, BPR & ``Ok, more similar voices" \\
User-to-Item & Personalization & BPR & \gray{user\_id:10021}  \\ \midrule
\multicolumn{4}{l}{{\color[HTML]{9B9B9B} \textit{Generative Retrieval}}} \\
Semantic ID & Multimodal In-Context & Residual VQ Tables & \gray{audio:semantic\_id:[63, 36, 44, 3]} \\ \bottomrule
\end{tabular}
}
\vspace{-4mm}
\end{table}

\textbf{SQL (Boolean Retrieval):} The SQL tool enables precise and structured queries on relational music metadata. We construct a single table with fields including title, artist, album, release date, tempo, key, and popularity. Tool arguments include query for SQL statements and topk for result limits, e.g., \texttt{sql(query="SELECT * FROM tracks WHERE date>=2020 ORDER BY tempo", topk=100)}.

\vspace{1mm} \textbf{BM25 (Sparse Retrieval):} This tool provides classic token-based text retrieval using the BM25 ranking function~\cite{robertson1999okapi}. BM25 offers lexical matching, making it particularly effective for text queries where typos are common and exact string matching is difficult. We construct five text corpora: title, artist, album, lyrics, and attributes (semantic tags). Tool arguments include query, corpus, and topk, e.g., \texttt{bm25(query="taylor swift songs", corpus="artist", topk=100)}.

\vspace{1mm} \textbf{Text-to-Item (Dense Retrieval):} This tool enables semantic music discovery through natural language descriptions by mapping text queries to musical content across multiple modalities. Utilizing pretrained text~\cite{yang2025qwen3} and multimodal encoders~\cite{wu2023large,tschannen2025siglip}, the tool retrieves the most similar items based on cosine similarity in the corresponding embedding space. Tool arguments include query, item modality type, vector database type, and topk, e.g., \texttt{text\_to\_item(query="Tracks with an album cover that shows a baby swimming underwater.", item\_modality="image", vector\_db="image", topk=20)}.

\vspace{1mm} \textbf{Item-to-Item (Dense Retrieval):} This tool supports example-based recommendation by finding similar items using dense representations. The LLM generates track IDs based on previously recommended tracks or user-provided examples in multi-turn conversations, then we look up the corresponding embeddings~\cite{wu2023large,tschannen2025siglip,rendle2012bpr} from pre-extracted vector databases to perform similarity-based retrieval. Tool arguments include track ID, item modality type, vector database type, and topk, e.g., \texttt{item\_to\_item(track\_id="22L7bfCiAkJo5xGSQgmiIO", item\_modality="audio", vector\_db="audio", topk=20)}.

\vspace{1mm}  \textbf{User-to-Item (Dense Retrieval):} This tool provides personalized recommendations using user embeddings trained through listening history~\cite{rendle2012bpr}. Unlike other tools activated by user queries, this tool is activated using the user ID from the user profile information. Tool arguments include user\_Id and topk, e.g., \texttt{user\_to\_item(user\_id=10021, topk=200)}.

\vspace{1mm} \textbf{Semantic IDs (Generative Retrieval): } Semantic IDs are discrete representations derived from item content features~\cite{rajput2023recommender,mei2025semantic}. Content is encoded into dense embeddings, then quantized into discrete codebook indices using Residual Vector Quantizer (RVQ). Semantic IDs serve as in-context information to help the LLM understand the multimodal properties of music. We build inverted indexes from code positions to item IDs for fast lookup by exact code match or small edit distance. Tool arguments include item modality type, Semantic ID indices, and topk, e.g., \texttt{semantic\_id(item\_modality="audio", indices=[52, 42, 5, 9], topk=20)}.

\subsection{In-Context Information for Inference}
In the proposed framework, the recommendation is made based on the in-context information provided to the LLM. This conditioning prompt includes four types of information

\textbf{System Prompt:}
The two stages of our framework have two distinct prompts: one for tool calling and, the other for response generation. For tool calling, we design a structured three-stage prompt for tool calling that decomposes the complex recommendation process into planning, retrieval, and reranking phases. Stage 1 (Planning) requires the LLM to select the exact retrieval tool and the reranking tool with a rationale for each choice. Stage 2 (Retrieval) executes the selected retrieval tool to gather at least \texttt{topk} unique \texttt{track\_id}s from the music database. Stage 3 (Reranking) applies the selected reranking tool to reorder the candidates to improve recommendation quality. The workflow enforces strict constraints: tools must be used in sequence (retrieval → reranking) with complementary roles and no functional overlap. The response generation prompt is designed to clearly explain the recommendation results, ensuring that the recommended track aligns with user queries.  

\textbf{Tool Functions:} We provide a comprehensive list of tool functions with their JSON schema that includes names, descriptions, and parameter data types. Detailed function examples are included in the Appendix~\ref{appx:1}.

\textbf{User Profiles:} For personalization, we provide user demographics and recent listening history as in-context information to the LLM. The demographics include User ID, type, age group, and gender. The User ID serves as in-context information for inference to properly activate the "User-to-Item" tool. For recent listening history, we include metadata from the five most recent tracks from the user's listening session, where each track's information includes metadata, attributes, and Semantic IDs.

\textbf{Previous Conversation History:} We provide the previous conversation state as in-context information to the LLM. The conversation state includes the user queries, the recommended musics, and the responses from the LLM. To incorporate multimodal information of music, we include not only track IDs but also metadata, attributes, and Semantic IDs as in-context information. 


\vspace{-3mm}
\section{Experiments}~\label{experiments}
\vspace{-7mm}

\textbf{Dataset:} We use TalkPlayData~2~\cite{choi2025talkplaydata}, a synthetic dataset designed for multimodal conversational music recommendation. The dataset contains diverse user profiles with demographic information (gender, age, country) from LFM-2b~\cite{schedl2022lfm}. It also provides comprehensive multimodal music representations including listening history, metadata, semantic tags, lyrics, album art, and audio. We evaluate different conversational recommendation systems on 1000 test conversations, each consisting of 8 turns.

\textbf{Environment:} Our environment consists of an executor and databases for tool calling. For SQL and BM25~\cite{robertson1999okapi}, we use basic metadata (track name, album name, artist name) and last.fm genre/style annotations from LFM-2b~\cite{schedl2022lfm}. For dense retrieval tools, we construct multiple vector databases using various pretrained models: Qwen3-0.6 embedding~\cite{yang2025qwen3} for text modality, CLAP~\cite{wu2023large} for audio modality, SigLIP2~\cite{tschannen2025siglip} for image modality, and Bayesian Personalized Ranking (BPR)~\cite{rendle2012bpr} for user and item embeddings. Finally, for Semantic IDs~\cite{rajput2023recommender, mei2025semantic}, we train a separate Residual Vector Quantizer (RVQ) on pre-extracted each representations (listening history, metadata, semantic tags, lyrics, album art, and audio), and then combine them to generate multimodal semantic identifiers. The RVQ architecture employs 4 residual quantization layers with 64 codebooks per layer, optimized with commitment loss. The hyperparameters were selected through validation experiments designed to balance the commitment loss and codebook utilization. To ensure fair evaluation and prevent data leakage, we employ chronological data splits for training both the BPR and RVQ models, where trainset strictly precedes testset.

\textbf{Evaluation:} For conversational recommendation, we measure Hit@K as our primary metrics with $k$=\{1,10,20\}. We compare our tool-calling approach against several baseline methods: (1) BM25, a classic sparse retrieval method; (2) Qwen3-LM + BM25, which combines language model generation with fuzzy matching; and (3) our proposed Qwen3-LM + Tool Calling approach that leverages multiple tools. We use Qwen3-LM-4B~\cite{yang2025qwen3} as our base model,~\footnote{https://huggingface.co/Qwen/Qwen3-4B} leveraging its strong reasoning capabilities for complex tool calling and query decomposition. We set the generation parameters to temperature=0.6 and top\_p=0.95. We represent all identifiers (User IDs, Track IDs, Semantic IDs) as natural language strings rather than special tokens with vocabulary expansion.

\section{Results}
\label{results}

\begin{table}[!t]
\caption{Conversational music recommendation results. QU, RG, TC indicate query understanding, response generation, tool calling capabilities, respectively.}
\label{tab:results}
\centering
\resizebox{\columnwidth}{!}{
\begin{tabular}{llcccrrr}
\toprule
Models & Methods & QU & RG & TC & Hit@1($\uparrow$) & Hit@10($\uparrow$) & Hit@20($\uparrow$) \\ \midrule
BM25 & Sparse & \cmark & \xmark & \xmark & 0.017 & 0.073 & 0.107 \\
Qwen3-LM + BM25 & Generative & \cmark & \cmark & \xmark  & 0.018 & 0.076 & 0.110 \\
Qwen3-LM + Tool (\textbf{Ours}) & Generative & \cmark & \cmark & \cmark  & 0.022 & 0.082 & 0.111
\\ \bottomrule
\end{tabular}
}
\vspace{-5mm}
\end{table}

\begin{figure}[t]
\centering
\includegraphics[width=\columnwidth]{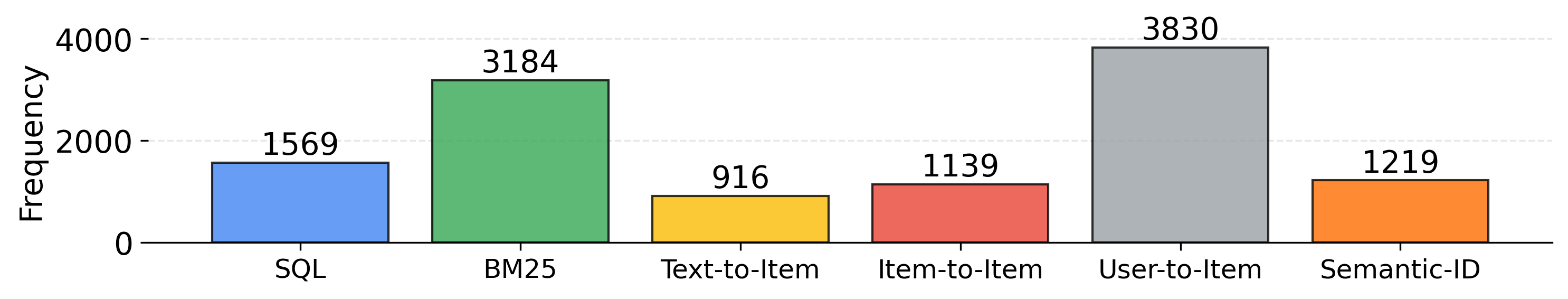}
\vspace{-6mm}
\caption{Tool Calling Frequency at First Attempt.}
\label{fig:tool_frequency}
\includegraphics[width=\columnwidth]{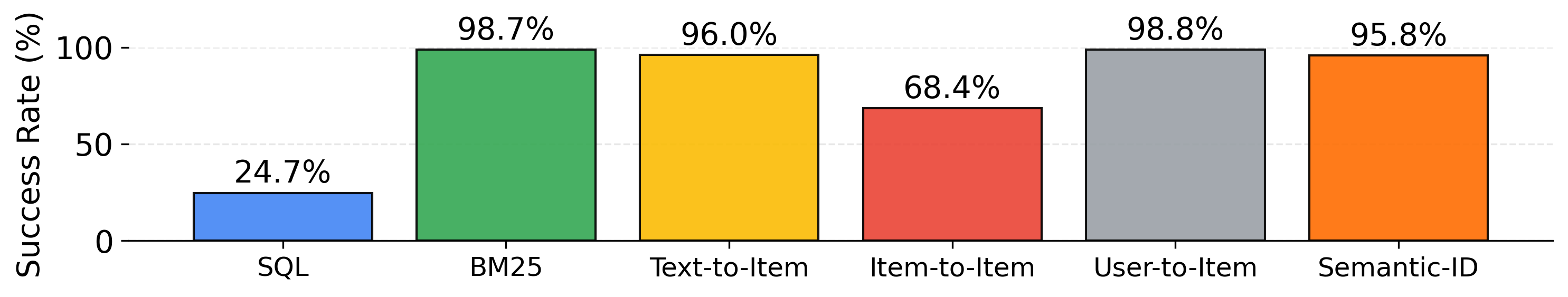}
\vspace{-6mm}
\caption{Tool Calling Success Rate at First Attempt.}
\label{fig:tool_success}
\vspace{-6mm}
\end{figure}

\textbf{Multi-turn Recommendation:} Table~\ref{tab:results} demonstrates that our tool-calling approach outperforms baseline methods in zero-shot conversational music recommendation. The improvement in Hit@1 performance (0.022 vs 0.018 for Qwen3-LM + BM25) highlights the effectiveness of reranking through multiple tool integration. When a tool call once fails during inference (as discussed later in detail), our system automatically retries the call, ensuring that all inference steps are completed successfully.

\textbf{Tool Calling Frequency:} Figures~\ref{fig:tool_frequency} and~\ref{fig:tool_success} analyze tool calling patterns based on the first attempt before any retry mechanisms. After repeated retries, in our experiment, the system always reaches to extract a result. The tool frequency distribution reveals that natural language-friendly tools like SQL and BM25 show higher usage frequencies, while item-specific modalities such as item-to-item matching and semantic ID tools exhibit lower frequencies. This pattern likely reflects pretraining exposure: universal operators such as SQL and BM25 commonly appear across information retrieval corpora, whereas item-to-item matching and Semantic ID tools tend to have domain-specific function names and interfaces. Consequently, pretrained LLMs are less familiar with these tools and invoke them less frequently.

\textbf{Tool Calling Success Rate:} However, the success rate analysis reveals performance challenges with more complex tools. SQL queries achieve only a 27.4\% success rate due to their syntactic complexity and common errors such as incorrect SQL syntax, using invalid column names, and retrieval failures caused by synonyms or typos leading to metadata-query mismatches. Item-to-item matching shows a 68.4\% success rate, which can be attributed to the challenge of predicting a complete track ID, which represents private information that is related to the music catalog. Unexpectedly, tools that rely on unseen information types, such as User IDs and Semantic IDs that were never encountered during LLM pretraining, demonstrate high success rates. User-to-Item personalization achieves a 98.8\% success rate, and Semantic ID retrieval shows a 95.8\% success rate. This high performance stems from the rich in-context information provided through user profiles and previous conversation history.

\begin{table}[!t]
\caption{
An inference example with the LLM \bluetext{inputs} and \orangetext{outputs}.
}
\resizebox{\columnwidth}{!}{
\begin{tabular}{l}
\toprule
\bluetext{Input: User Demograph} \\ 
UserID: N/A (Cold User), Age group: 20s, Country: Slovenia, Gender: Female, Previous Tracks: N/A (Cold User)\\ \midrule
\bluetext{Input: Chat Context (Previous Turns)} \\
\begin{tabular}[c]{@{}l@{}}
\textit{User}: I'm looking for a German electronic artist, I think they had an album called 'Walls'\\with some really minimal and melancholic tracks.\end{tabular} \vspace{1.5mm} 
\\
\begin{tabular}[c]{@{}l@{}} \textit{Assistant}: TrackID: 1KsqDnRQXFdFypQdFkB0wA,\\ 
title: not a number, artist: apparat, album: walls, tags: experimental, 00s, post-rock, instrumental, soundtrack \\ 
ambient, electronica, electronic, tempo: 130.37, key: C major, release\_date: 2007.05.25 \\ 
`audio:semanticID': [0, 39, 63, 53], `image:semanticID': [41, 10, 17, 48], `metadata:semanticID': [15, 49, 15, 36],\\
`lyrics:semanticID': [45, 38, 39, 50], `attributes:semanticID': [20, 55, 58, 51], `cf\_item:semanticID': [59, 21, 42, 34]\}
\end{tabular} \vspace{1.5mm} 
\\
\begin{tabular}[c]{@{}l@{}} \textit{Assistant}: Oh, absolutely! Apparat from the Walls album is exactly what I was thinking.\\For a really minimal and melancholic vibe, you should definitely check out Not A Number.\\ It perfectly captures that atmospheric, introspective feeling.
\end{tabular} \vspace{1.5mm} 
\\
\begin{tabular}[c]{@{}l@{}} \textit{User}: Yes, this is exactly the artist I was looking for! Perfect, Apparat is the one.\\Can you recommend some other tracks from their discography, 
especially more instrumental or ambient ones? \end{tabular} 
\\ \midrule
\orangetext{Intermediate: Chain-of-thought for Tool Calling} \\
\begin{tabular}[c]{@{}l@{}}
Okay, the user just confirmed that Apparat is the artist they were looking for and wants more recommendations \\ 
from their discography, specifically instrumental or ambient tracks. Let me break down how to approach this. \\
First, the user is a cold\_start user, so I need to use the appropriate tools.
\\
\gray{...[Omitted]...}
\\
Alternatively, using the 'artist' corpus\_type with BM25 to find tracks by Apparat, then filter for instrumental,ambient. \\
But the tools available include BM25, which is for lexical matching. Alternatively, using text\_to\_item with \\
modality\_type 'text' and corpus\_type 'attributes', querying for instrumental, ambient. 
\\
\gray{...[Omitted]...}
\\
So the first tool call would be BM25 with query 'Apparat', corpus\_type 'artist', topk 20. \\
Then, the second tool call would be text\_to\_item\_similarity with query 'instrumental, ambient',\\
modality\_type 'text', corpus\_type 'attributes', topk 20. But the user is looking for tracks from Apparat, \\
so the initial retrieval is tracks by Apparat, and the reranking is based on attributes. 
\end{tabular} \\ \midrule
\orangetext{Intermediate: Tool Calling Results} \\
\begin{tabular}[c]{@{}l@{}}
[\{"tool\_name": "bm25","tool\_args": \{"query": "Apparat",corpus\_type": "artist","topk": 20\},
\\ \hspace{0.7mm} \{"tool\_name": "text\_to\_item","tool\_args": \{"query": "instrumental, ambient","modality\_type": "text",
\\ \hspace{2mm}  "corpus\_type": "attributes","topk": 20\}]
\end{tabular} 
\\ \midrule
\orangetext{Output: Recommendation Track} \\
\begin{tabular}[c]{@{}l@{}}\textit{Assistant}: TrackID: 40HHuqH12ssXNpnHkiz0fk,\\title: tod, artist: apparat, album: krieg und frieden (music for theatre) \\ tags: experimental, ambient, dance, electronic, soundtrack, tempo: 85.83, key: A major, release\_date: 2013.02.19
\end{tabular} \\ \midrule
\orangetext{Intermediate: Chain-of-thoguht for Response} \\
\begin{tabular}[c]{@{}l@{}}
Okay, the user just confirmed that Apparat is the artist they were looking for and asked for more recommendations,\\
specifically instrumental or ambient tracks. The track provided is Tod from Krieg und Frieden (Music for Theatre).\\ 
I need to verify if Tod is indeed instrumental or ambient. The tags include experimental, electronic, and soundtrack. \\
The tempo is 85.83, which is slow, and the key is A major. The release date is 2013. 
\\
\gray{...[Omitted]...}
\\
I should present this track as a recommendation, highlighting the genre, mood, and why it's a good match. \\
I need to make sure the response is enthusiastic and confident, confirming that this track matches their preferences. \\
Then, invite them to explore more similar tracks or discuss other preferences.\\ 
\end{tabular} \\ \midrule
\orangetext{Output: Response} \\
\begin{tabular}[c]{@{}l@{}}
\textit{Assistant}: Absolutely! Tod from Krieg und Frieden (Music for Theatre) is a stunning ambient track that perfectly \\
complements the minimal, introspective vibe you love. It’s a slow, atmospheric piece with a haunting, melancholic \\ that
feels both electronic and deeply emotional.  The A major key and steady tempo create a meditative, \\
immersive experience. Would you like more recommendations or explore other moods?
\end{tabular}
\\ \bottomrule
\end{tabular}
}
\label{tab:inference_example}
\end{table}

\textbf{Qualitative Example}
Table~\ref{tab:inference_example} shows an inference example of TalkPlay-Tools. The inference process begins with user information and conversational chat context, including previously recommended tracks and user profile data. The system proceeds through a tool calling stage where the model analyzes the user query and identifies appropriate retrieval tools. The retrieved results are then used to generate natural language response. This multi-stage process ensures both retrieval accuracy and conversational fluency in the final response.

\section{Conclusion}
\label{conclusion}
We presented a novel tool-calling framework for conversational music recommendation that seamlessly integrates multiple modalities and the various components in a retrieval-reranking pipeline. For successful integration, we proposed orchestrating external executors and models through LLM tool calling. Using various types of tools, the proposed framework effectively harnessed multimodal information spanning text, audio, visual, and behavioral data to deliver enhanced music recommendations within conversational contexts. Future research directions present several promising avenues: (1) incorporating reinforcement learning methodologies to optimize tool calling precision and minimize reliance on retry mechanisms; (2) designing personalized tool calling strategies that extend beyond track-centric approaches to capture nuanced user preferences and behavioral patterns more effectively.

{\small
\bibliography{references}
\bibliographystyle{abbrv}
}


\appendix
\newpage

\section{Tool Calling Functions} \label{appx:1}
This appendix specifies the callable tools used by the agent in a Pythonic format. Following previous works~\cite{schick2023toolformer, doh2025llm2fx}, each block presents the function signature, purpose, input arguments, and expected return. All tools return a list of \texttt{track\_id} strings (up to \texttt{topk}), and are designed to be composed in sequence within the retrieval–reranking pipeline.

\begin{lstlisting}[language=Python]
def sql(sql_query: str, topk: int) -> list[str]:
    """
    Execute an SQL query for boolean (structured) matching.
    Must return results with track_id 
    (e.g., `SELECT track_id FROM tracks WHERE ...`).
    SQL schema (table: tracks):
        track_id TEXT PRIMARY KEY
        title TEXT
        artist TEXT
        album TEXT
        popularity INTEGER
        release_date DATE (YYYY-MM-DD)
        tempo REAL
        key TEXT

    Args:
        sql_query: SQL query string to execute.
        topk: Maximum number of track_ids to return.

    Returns:
        List[str]: Up to `topk` track_ids.
    """
\end{lstlisting}

\begin{lstlisting}[language=Python]
def bm25(query: str, corpus_type: str, topk: int) -> list[str]:
    """
    Perform BM25 retrieval for lexical matching.
    Lowercase all input strings.

    BM25 corpora:
        'title'      : lowercase of track name
        'artist'     : lowercase of artist name
        'album'      : lowercase of album name
        'lyrics'     : lowercase of lyrics
        'attributes' : lowercase of genre, instrument, mood, theme, usage, etc.

    Args:
        query: Search query string.
        corpus_type: 
        One of {"title", "artist", "album", "lyrics", "attributes"}.
        topk: Maximum number of track_ids to return.

    Returns:
        List[str]: Up to `topk` track_ids.
    """
\end{lstlisting}

\begin{lstlisting}[language=Python]
def text_to_item_similarity(
        query: str, modality_type: str,
        vector_db_type: str, topk: int
        ) -> list[str]:
    """
    Perform text-to-item semantic similarity retrieval.

    Args:
        query: Search query string.
        modality_type: One of {"text", "audio", "image"}.
        vector_db_type: One of {"metadata", "lyrics",
        "attributes", "audio", "image"}.
        topk: Maximum number of track_ids to return.

    Returns:
        List[str]: Up to `topk` track_ids (most similar to the text query).
    """
\end{lstlisting}

\begin{lstlisting}[language=Python]
def item_to_item_similarity(
        track_id: str, modality_type: str,
        vector_db_type: str, topk: int
        ) -> list[str]:
    """
    Perform item-to-item similarity retrieval (example-based recommendation).
    Note: `track_id` is a 22-character string.

    Args:
        track_id: Unique track identifier.
        modality_type: One of {"audio", "image", "cf"}.
        vector_db_type: One of {"audio", "image", "cf"}.
        topk: Maximum number of track_ids to return.

    Returns:
        List[str]: Up to `topk` track_ids similar to the input item.
    """
\end{lstlisting}

\begin{lstlisting}[language=Python]
def user_to_item_similarity(user_id: str, topk: int) -> list[str]:
    """
    Perform user-to-item similarity retrieval (personalization).
    Use only the `user_id` from demographic/profile info.
    If `user_type` is "cold_start", do not select this tool.

    Args:
        user_id: Unique user identifier (string).
        topk: Maximum number of results to return.

    Returns:
        List[str]: Up to `topk` personalized track_ids.
    """
\end{lstlisting}

\begin{lstlisting}[language=Python]
def semantic_id_matching(
        modality_type: str, 
        indices: list[int],
        topk: int
        ) -> list[str]:
    """
    Perform Semantic ID matching via codebook index lookups.

    Args:
        modality_type: One of {"audio", "image", "metadata", "lyrics", "attributes", "cf_item"}.
        indices: List of code indices (e.g., Residual VQ codes).
        topk: Maximum number of results to return.

    Returns:
        List[str]: Up to `topk` track_ids matched by Semantic IDs.
    """
\end{lstlisting}



\end{document}